# SCAN: An Efficient Density Functional Yielding Accurate Structures and Energies of Diversely-Bonded Materials

Jianwei Sun[1*], Richard C. Remsing[2,3], Yubo Zhang[1], Zhaoru Sun[1], Adrienn Ruzsinszky[1], Haowei Peng[1], Zenghui Yang[1], Arpita Paul[4], Umesh Waghmare[4], Xifan Wu[1], Michael L. Klein,[1,2,3] and John P. Perdew[1,2]

1. Department of Physics, Temple University, 1925 N. 12th St., Philadelphia, PA 19122, USA

2. Department of Chemistry, Temple University, 1901 N. 13th St., Philadelphia, PA 19122, USA

3. Institute for Computational Molecular Science, Temple University, 1925 N. 12th St., Philadelphia, PA 19122, USA

4. Theoretical Sciences Unit, Jawaharlal Nehru Centre for Advanced Scientific Research, Bangalore 560064, India

**Kohn-Sham density functional theory[1] (DFT) is a widely-used electronic structure theory for materials as well as molecules. DFT is needed especially for large systems, *ab initio* molecular dynamics, and high-throughput searches for functional materials. DFT's accuracy and computational efficiency are limited by the approximation to its exchange-correlation energy $E_{xc}$. Currently, the local density approximation (LDA)[1,2] and generalized gradient approximations (GGAs)[3] dominate materials computation mainly due to their efficiency. We show here that the recently developed non-empirical strongly constrained and appropriately normed (SCAN) meta-GGA[4] improves significantly over LDA and the standard Perdew-Burke-Ernzerhof GGA[3] for geometries and energies of diversely-bonded materials (including covalent, metallic, ionic, hydrogen, and van der Waals bonds) at comparable efficiency. Thus SCAN may be useful even for soft matter. Often SCAN matches or improves upon the accuracy of a computationally expensive hybrid functional, at almost-GGA cost. SCAN is therefore expected to have a broad impact on materials science.**

LDA, the earliest approximation in DFT, constructs a local energy density from just the local electron density, and is exact for any uniform electron gas[1]. It usually overestimates the strengths of all bonds near equilibrium. By building in the electron density gradient, the standard Perdew-Burke-Ernzerhof (PBE) GGA[3] softens the bonds to give robust and overall more accurate descriptions, except for the van der Waals (vdW) interaction which is largely lost. By mixing

GGAs with the nonlocal exact exchange, hybrid GGAs (e.g., the PBE0 hybrid GGA[5] where 25% of the exact exchange energy is mixed with 75% of PBE GGA exchange) can improve further the descriptions for the covalent, ionic, and hydrogen bonds. However, hybrid GGAs still fail to describe the vdW interaction, and the inclusion of the nonlocal exact exchange deteriorates the computational efficiency dramatically, especially so for metallic systems. The computational cost of a hybrid functional can be 10 to 100 times[6] that of a semilocal functional in standard codes. Another problem with hybrids is that a universal exact-exchange mixing parameter is not determined by any exact condition. On the other hand, compared to GGA, meta-GGA includes the electron kinetic energy density as an input to model $E_{xc}$ and therefore is still semilocal and efficient in computation. The inclusion enables meta-GGAs to recognize and accordingly treat different chemical bonds (e.g., covalent, metallic, and even weak bonds), which no LDA or GGA can[7]. The SCAN meta-GGA[4] satisfies all 17 exact constraints that a meta-GGA can, and is appropriately normed on systems for which semilocal functionals can be exact or nearly-exact. SCAN uses no bonded information in its construction, and thus is nonempirical with genuine predictive power.

*vdW bonding in ice phases:* It was believed that conventional semilocal and hybrid functionals were incapable of describing vdW bonds that arise from intermediate-range vdW interactions. vdW interactions are typically weak, but still important, e.g., for the structures of a hydrogen-bonded network like ice. In the binding energy difference per $H_2O$ between one ice phase and another, any errors in the hydrogen bonding energies tend to cancel out, but errors in the vdW bonding energies do not, because the vdW attraction increases with the density of water molecules. Figure 1 shows that both PBE and PBE0 significantly destabilize high-pressure phases relative to Ih (the stable phase of ice at ambient pressure), while the addition of the Tkatchenko-Scheffler[8] vdW correction (vdW_TS) improves the energy differences dramatically compared to the experimental results or the highly accurate yet expensive diffusion Monte Carlo (DMC) predictions[9,10]. Unfortunately, both PBE+vdW_TS and PBE0+vdW_TS miss one fine detail: they predict that ice IX is more stable than ice II, whereas ice II was determined to be 3 meV/ $H_2O$ more stable than ice IX in experiments. The small difference in energy between phases IX and II is likely due to many-body contributions to the dispersion interactions in these systems, and the pair-wise vdW_TS correction does not account for such effects. Interestingly and surprisingly, the SCAN meta-GGA[4] predicts that ice II is 3 meV/ $H_2O$ more stable than ice IX, in complete agreement with experiment. Moreover, SCAN yields energy differences between all the different ice phases studied here with accuracy comparable to that of PBE0+vdW_TS. SCAN even considerably improves upon the predictions of PBE0+vdW_TS for the energy difference between ice Ih and the high-density phase VIII. The lower panel of Fig. 1 also shows that SCAN predicts volume changes between ice phases in near-quantitative agreement with experimental results, and thus with greater accuracy than all other functionals considered here. The performance of SCAN on the ice polymorphs defies the conventional wisdom and shows that SCAN has the ability to capture the intermediate-range vdW interaction[10,11]. Of course, SCAN cannot describe the long-range vdW interaction that exists even for non-overlapping electron densities.

*Covalent and hydrogen bonding in isolated water monomer, dimer, and ice Ih:* In $H_2O$, the molecule that is indispensable to life, the hydrogen atoms and the oxygen atom are covalently bonded with an angle of about 105 degrees between the OH bonds. All the functionals considered in Table I give reasonably accurate predictions for the bond length and the bond angle as well as the vibrational frequencies and the dipole moment of a single water molecule, while PBE0 and SCAN are the best among them in comparison with the experimental results, demonstrating that SCAN is highly accurate for covalent bonds.

Due to its unacceptable overestimation of the strength of hydrogen bonds, LDA is seldom used for the study of water. PBE, PBE0, and SCAN, as shown in Table I, all predict satisfactorily accurate properties for the water dimer and ice Ih, where a good description for the hydrogen bond is vital. PBE slightly overestimates the binding energies of the water dimer and the lattice energy of ice Ih. PBE0 improves over PBE after including a portion of the exact exchange. Although SCAN is comparable to PBE0 for the properties of the water monomer, it overestimates slightly more than PBE the binding energies of the water dimer and the lattice energy of ice Ih.

*Covalent and metallic bonding in crystalline and liquid Si:* Silicon, the base material for the semiconductor industry, crystallizes at ambient conditions in the diamond structure with a coordination number of 4, and undergoes a semiconductor-metal phase transition around 12 GPa of pressure into the $\beta$-Sn structure with a coordination number of 6. Fig. 2 (a) shows that, compared to experiment or to computationally-expensive high-level diffusion Monte Carlo (DMC) calculations[12], both LDA and PBE give accurate volumes for Si in the diamond and $\beta$-Sn phases, but PBE has a more realistic yet still unsatisfactory energy difference between them. On the other hand, SCAN and the Heyd-Scuseria-Ernzerhof (HSE) hybrid GGA[13] predict the energy difference of the two phases in excellent agreement with the DMC result. HSE is a range-separated version of the PBE0 hybrid GGA, which replaces the long-range part of the exact exchange with that of the PBE GGA to improve the efficiency and to alleviate the problematic behavior in metallic systems.

It has been argued[12] that semilocal functionals do not predict an accurate energy difference between these two phases because they underestimate the band gap of the diamond phase significantly, while HSE improves the band gap and therefore the energy difference. However, this is not a satisfactory explanation. SCAN gives a band gap of diamond Si only about halfway between PBE and experiment, but a very accurate structural energy difference. The improvement of the energy difference comes from the ability of SCAN to distinguish between covalent and metallic bonds and to properly stabilize covalent single bonds[14].

Fig. 2 (b) shows that SCAN significantly improves over LDA and PBE for the interstitial defect formation energies of diamond Si, reaching the level of accuracy of HSE. The three lowest-energy interstitial defects in Si are T (tetrahedral site), H (hexagonal), and X (split), for which DMC predicts defect formation energies of 5.05, 5.13, and 4.94 eV, respectively[15]. However, these DMC values were calculated by using GGA-relaxed geometric structures in a small supercell. The best interstitial formation energies estimated from experiments[12] are in the range of 4.23-4.85 eV, with which the results of both SCAN and HSE are in excellent agreement.

Upon melting, silicon undergoes a transition from a semiconducting solid to a metallic liquid that contains transient covalent bonds between neighboring atoms. The properties of liquid Si (*l*-Si) depend sensitively on the relative amounts of metallic and covalent bonding present in solution, and we find that SCAN provides a good description of this complex liquid. Simulations of *l*-Si in the isothermal-isobaric ensemble at T=1800K and P=0 bar yield a density of 2.57 g/cm$^3$ from SCAN, in good agreement with the experimental value[16] of 2.59 g/cm$^3$, while PBE (2.54 g/cm$^3$) slightly underestimates and LDA (2.70 g/cm$^3$) overestimates the density as expected. SCAN also yields a position of the first peak of the pair correlation function *g(r)* in excellent agreement with that of experiment[17], as shown in Fig 3 (b), while those of LDA and PBE are shifted to slightly larger distances. The SCAN description of *l*-Si leads to a pronounced second peak in *g(r),* as in the experimental results, albeit shifted to larger distances. Such a pronounced second peak is lacking in both LDA and PBE descriptions. The increased accuracy of SCAN with respect to PBE and LDA is again due to better discrimination of metallic and covalent bonds, with the latter manifesting the tetrahedral coordination structure of molten Si, highlighted by the simulation snapshot showing electron density corresponding to covalent bonds between silicon atoms in a tetrahedral arrangement.

*Ionic bonding in ferroelectric and multiferroic materials:* Interactions between ionic species are primarily electrostatic in origin, but can also have a significant component of vdW interactions among highly-polarizable negative ions, for example, making the description of such systems challenging. These situations often arise in ferroelectric materials like the prototypical $BaTiO_3$ and $PbTiO_3$, which exhibit spontaneous electric polarization due to structural instabilities at low temperature[18], and $BiFeO_3$, a multiferroic material with ferroelectric and antiferromagnetic properties[19]. The prediction of structural instabilities from first-principles calculations is extremely sensitive to volume changes, and even small errors of 1-2% in lattice constants obtained from LDA and PBE yield unsatisfactory predictions for ferroelectric materials. PBE, for example, is particularly poor in its description of these materials, as it predicts spurious supertetragonality (too large c/a) in $BaTiO_3$ and $PbTiO_3$[18].

There have been efforts to design functionals for solids to remedy this deficiency. The B1WC hybrid GGA[18] was designed for ferroelectric materials. It mixes 16% of exact exchange energy with 84% of Wu-Cohen GGA exchange[20] to optimize the properties of $BaTiO_3$. Table II shows that B1WC predicts volumes for these three materials in excellent agreement with the experimental results, and also very accurate c/a ratios and polarizations for $BaTiO_3$ and $PbTiO_3$. On the other hand, the more commonly used HSE hybrid GGA inherits the spurious supertetragonality for BaTiO3 and PbTiO3 from its parent PBE GGA[18] (although less severe), and predicts too large polarizations. SCAN is overall almost comparable to the computationally expensive B1WC, and much better than LDA and PBE for the above properties. The SCAN energy differences between the cubic and tetragonal phases are much closer to the B1WC values than either LDA, PBE, or even HSE.

SCAN also gives more realistic band gaps for these compounds than LDA and PBE, consistent with our findings in Si and other semiconductors. This is possible because the SCAN meta-GGA, like the hybrid functionals, is implemented in a generalized Kohn-Sham scheme in which the exchange-correlation potential is not a multiplicative operator. Hybrid gaps are however more realistic than SCAN gaps.

For the magnetic moment of Fe in $BiFeO_3$, PBE predicts the most accurate value (3.70 µB) in comparison with the experimental one (3.75 µB), while SCAN is the second best with 3.96 µB and B1WC significantly overestimates this value. Remarkably, for ferroelectrics and multiferroics the nonempirical and semilocal SCAN meta-GGA is often comparable to or better than a hybrid functional fitted to $BaTiO_3$.

In studies of multiferroics, where late *3d* transition metals are usually present to provide the ferromagnetic properties, the Hubbard U is introduced for LDA and PBE to account for the on-site Coulomb interaction, and thus to open the band gap. Table II shows that the SCAN band gap is comparable to that of PBE+U with U=2 eV for the Fe atoms[21]. Both SCAN and PBE+U give similar magnetic moments for Fe and comparable polarizations. However, SCAN is much better for the volume.

*Summary:* We have demonstrated that accurate structures and energies of materials with diverse bonding are predicted by the nonempirical SCAN meta-GGA. Without being fitted to any bonded system, SCAN accurately describes all kinds of bonding. These successes were unexpected from a computationally-efficient functional. The examples unambiguously show that SCAN is more accurate than LDA and GGA but with comparable efficiency, and is often as or more accurate in comparison with hybrid GGAs.

**Method**

All our DFT calculations are self-consistent. Most of the *ab initio* calculations for the water monomer and dimer were carried out in the GAUSSIAN[22] code, except for those of PBE+vdW_TS and PBE0+vdW_TS which were performed in FHI-aims[23]. The geometric, vibrational, and electrostatic properties were calculated with the aug-cc-pvtz basis set in GAUSSIAN and the tier-3 basis set in FHI-aims. The binding energies of the water dimer were obtained by extrapolating to the complete basis set limit. The calculations for solids and liquids were performed using the VASP code and PAW potentials in the implementation of Kresse and Joubert[24]. For ice polymorphs, we used the geometries and the computational settings of Ref. 8. The phase-transition calculations for silicon followed the settings of Ref. 14. The interstitial defect calculations used an energy cutoff of 400 eV and a gamma-centered $4 \times 4 \times 4$ k-mesh. The defects were placed in and relaxed with the host atoms of a 64-atom simulation cell, with the lattice constant determined by the underlying functionals. AIMD simulations of liquid Si were performed with simulation cells of 216 atoms. An energy cutoff of 300 eV was used for the silicon AIMD calculation. 20 ps production runs and the gamma-only k-mesh were used in all AIMD calculations and analysis. For simulations of ferroelectric and multiferroic materials, an energy cutoff of 600 eV was used. We used a tetragonal cell of 5 atoms and a gamma-centered $8 \times 8 \times 8$ k-mesh for $BaTiO_3$ and $PbTiO_3$, and a hexagonal cell of 30 atoms and a gamma-centered $4 \times 4 \times 2$ k-mesh for $BiFeO_3$. The spin configuration of $BiFeO_3$ was fixed to the G-type antiferromagnetic state. The spin-orbit coupling effect was neglected for all calculations. The spontaneous polarization was calculated according to the modern theory of polarization[25].

**Acknowledgements:** This research was supported as part of the Center for the Computational Design of Functional Layered Materials, an Energy Frontier Research Center funded by the U.S. Department of Energy (DOE), Office of Science, Basic Energy Sciences (BES), under Award # DE-SC0012575. Computer equipment in Temple's HPC Center was supported by the National Science Foundation (NSF) under major research instrumentation grant number CNS-09-58854. J.S., R.C.R., Y.Z, Z.S., A.R., and H.P. thank National Energy Research Scientific Computing Center (NERSC), a DOE Office of Science User Facility, and the HPC center of Temple University for computer time. J.S., X.W., and J.P.P. thank R. Car, B. Santra and R. DiStasio Jr. for helpful discussions.

**Author Contributions:** J.S. and J.P.P. designed the project. J.S., R.C.R., Y. Z, Z.S., A.R., and H.P. carried out the calculations. J.S. implemented the SCAN metaGGA and prepared the initial manuscript. All authors contributed to the discussions and revisions of the manuscript.

**Additional information:** The authors declare no competing financial interests. Correspondence and requests for materials should be addressed to J.S.

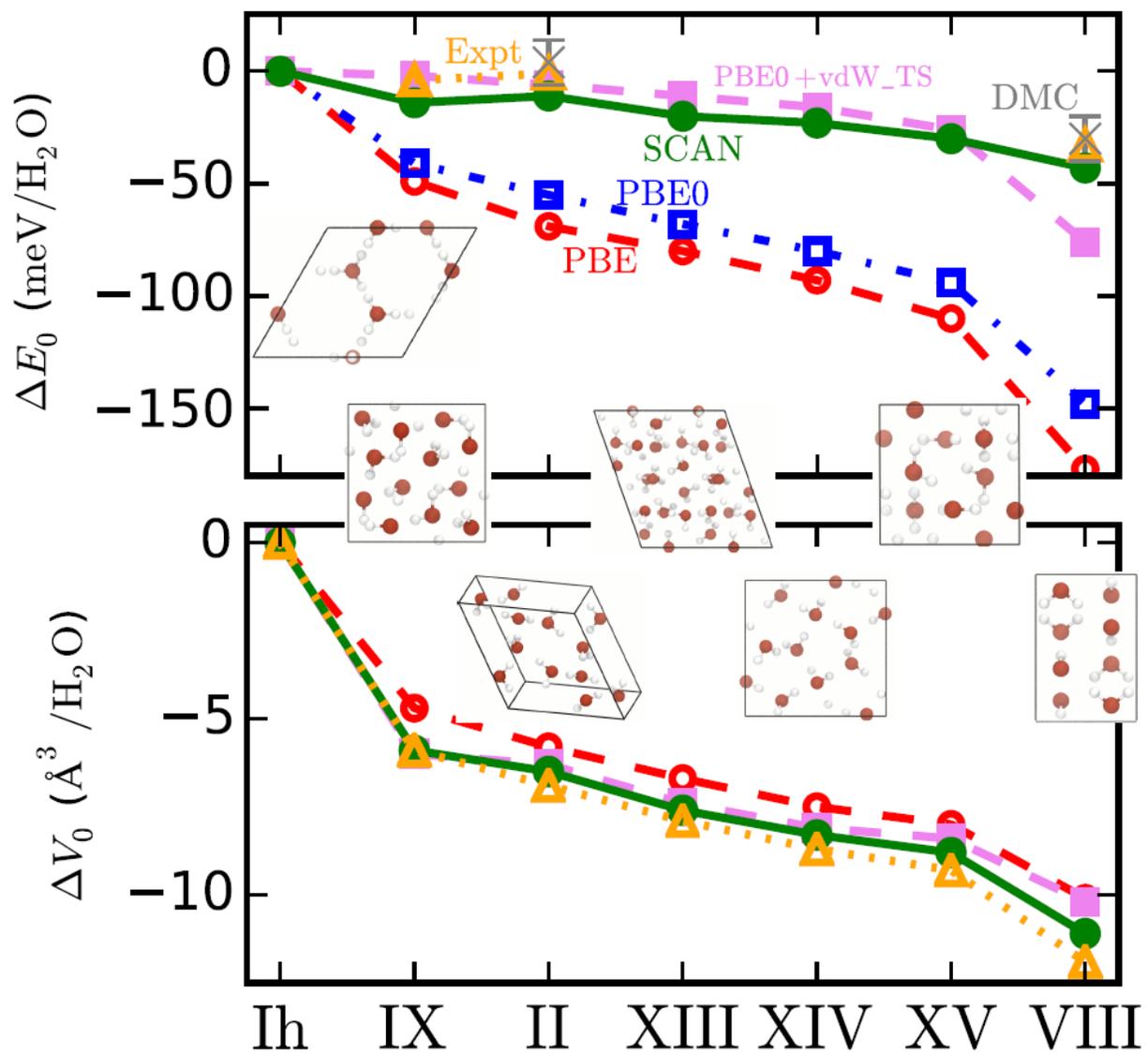

Fig 1. Relative binding energy and equilibrium volume changes per $H_2O$ unit of 7 hydrogen-ordered ice phases with respect to the ground state ice Ih. The zero-point energy effects have been extracted in the experimental results[26,27]. The values other than those of SCAN are from Refs. 9.

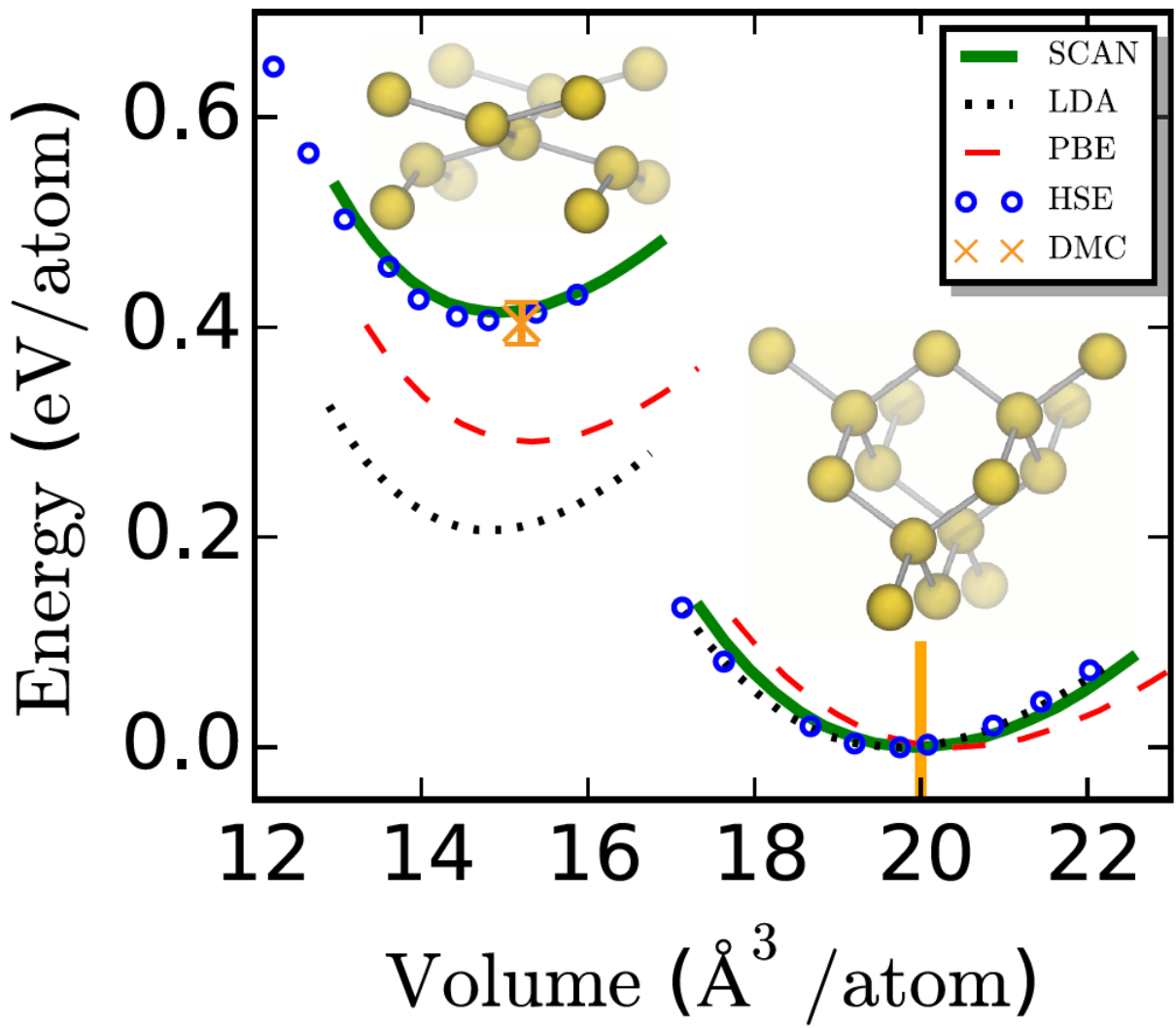

(a)

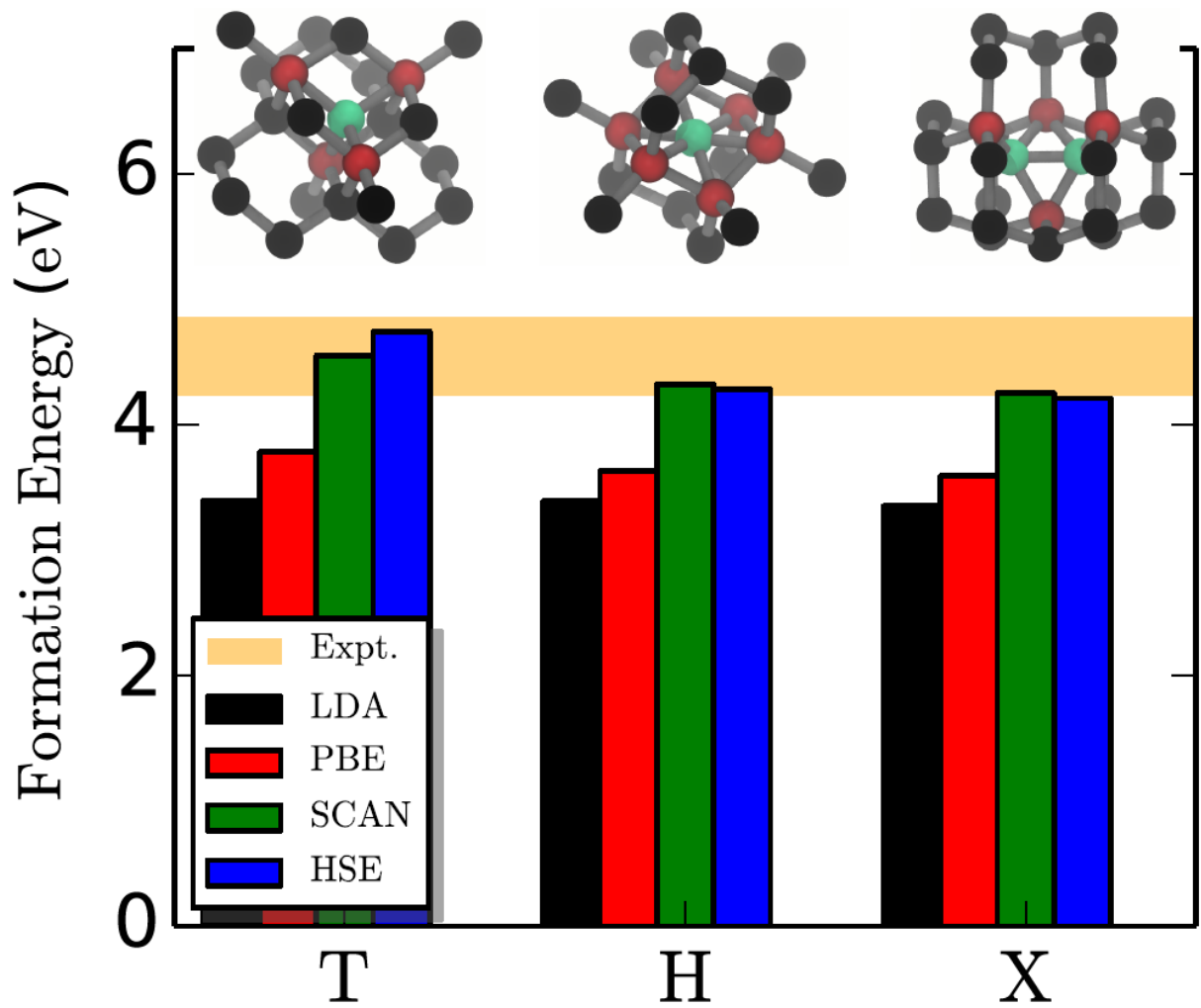

(b)

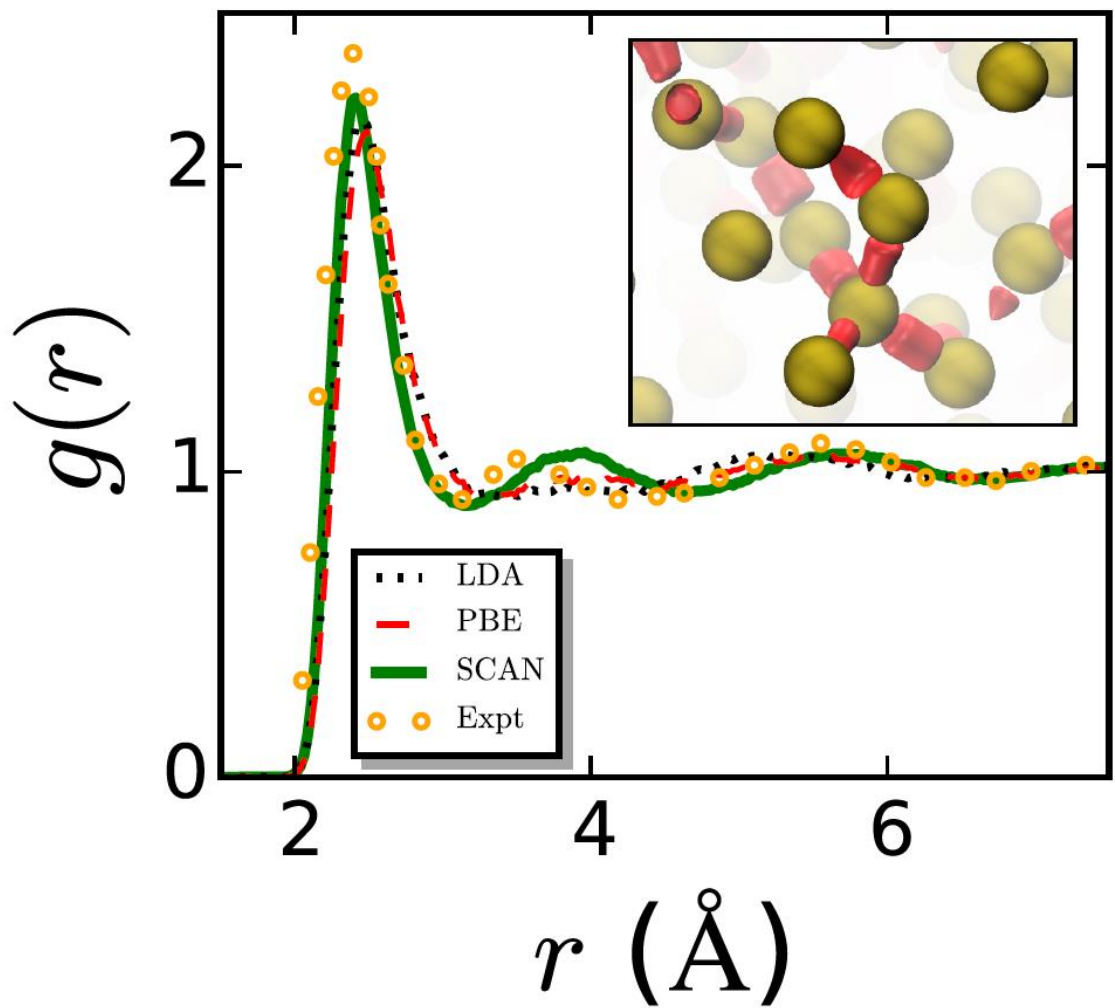

(c)

Fig. 2. (a) Energy difference between $\beta$-Sn and diamond phases of Si. The orange vertical line indicates the experimental volume[12] of diamond Si. (b) Interstitial defect formation energy in diamond Si. Green atoms are the defects, red atoms the nearest-neighbors of the defects, and black atoms those in the second coordination shell. (c) Pair correlation function, *g(r),* of liquid Si at T=1800K, as well as a snapshot of the liquid Si simulation using SCAN, with the red isosurfaces of electron density corresponding to covalent bonds between silicon atoms (yellow) in a tetrahedral arrangement (the bottom right inset). The experimental *g(r)* is from Ref. 17.

Table I. Properties of the water monomer, water dimer, and ice Ih at equilibrium. For the water monomer, R(O-H) is the bond length (Å), ∠HOH the bond angle, ($\nu_1$, $\nu_2$, $\nu_3$) the vibrational frequencies (cm$^{-1}$) ($\nu_1$ and $\nu_2$ are the asymmetric and symmetric O-H stretching modes and $\nu_3$ is the H-O-H bending mode), $\mu$ the dipole moment (Debye), $\alpha$ the isotropic polarizability, and $\beta$ the magnitude of hyperpolarizability. For the water dimer, R(O-O) is the distance between the two oxygen atoms, and $E_b$ the binding energy (meV/ $H_2O$). The last two rows give volume $V$ (Å$^3$/$H_2O$) and lattice energy $E_0$ (meV/ $H_2O$) of ice Ih at equilibrium. The experimental values and the best *ab initio* estimations for the water monomer and dimer are from Ref 28, except for the best *ab initio* estimations of $\alpha$ and $\beta$, where the CCSD values from Ref 29 are used. For ice Ih, the zero-point energy effects have been extracted in the experimental results[26,27], and the values other than LDA and SCAN are from Ref. 9.

|  | LDA | PBE | PBE0 | PBE+vdW_TS | PBE0+vdW_TS | SCAN | Best *ab initio* | Expt. |
|---|---|---|---|---|---|---|---|---|
| R(O-H) | 0.970 | 0.970 | 0.959 | 0.969 | 0.957 | 0.961 | - | 0.957 |
| ∠HOH | 105.0 | 104.2 | 104.9 | 104.2 | 104.9 | 104.4 | - | 104.5 |
| $\nu_1$ | 3837 | 3802 | 3962 | 3810 | - | 3911 | - | 3943 |
| $\nu_2$ | 3726 | 3697 | 3857 | 3706 | - | 3806 | - | 3832 |
| $\nu_3$ | 1551 | 1592 | 1633 | 1595 | - | 1647 | - | 1648 |
| $\mu$ | 1.858 | 1.804 | 1.854 | 1.804 | 1.853 | 1.847 | - | 1.855 |
| $\alpha$ | 10.24 | 10.33 | 9.58 | - | - | 9.81 | 9.65 | 9.63±0.20 |
| $\beta$ | 36.0 | 35.5 | 28.7 | - | - | 34.4 | 29.6 | - |
| R(O-O) | 2.71 | 2.90 | 2.89 | 2.90 | 2.89 | 2.86 | 2.91 | 2.98 |
| $E_b$ | 193.6 | 109.7 | 106.7 | 117.8 | 113.7 | 118.3 | 108.8 | 118±15 |
| V | 25.37 | 30.79 | 30.98 | 29.67 | 29.88 | 29.56 | 31.69±0.01 | 30.91 |
| $E_0$ | 1095 | 636 | 598 | 714 | 672 | 660 | 605±5 | 610 |

Table II. Properties of prototypical ferroelectric (BaTiO$_3$ and PbTiO$_3$) and multiferroic (BiFeO$_3$) materials predicted by LDA, PBE with and without the Hubbard U correction, SCAN, and hybrid GGAs. In the PBE column, the values of PBE with U=2 correction are in parentheses. The B1WC hybrid GGA[18] designed for ferroelectric materials is used as the reference in the second last columns. Eg (eV) is the fundamental band gap, V (Å$^3$) the volume, c/a the ratio of the lattice constants c and a, and Ps ($C/m^2$) the polarization of the tetragonal phases. $\Delta E$(meV/cell) is the total energy difference between the cubic and tetragonal phases. $\mu$ ($\mu B$) is the magnetic moment per Fe. The B1WC and experimental results for BaTiO$_3$ and PbTiO$_3$ are from Ref. 18, while the hybrid GGAs and experimental results for BiFeO$_3$ are from Ref. 19. The experimental polarization for BiFeO$_3$ is taken from Ref. 30.

| systems | Property | LDA | PBE (U=2) | HSE | SCAN | B1WC | Expt. |
|---|---|---|---|---|---|---|---|
| BaTiO3 | Eg | 1.72 | 1.73 | 3.27 | 2.13 | 3.44 | 3.38 |
|  | V | 62.1 | 67.5 | 64.5 | 65.1 | 63.2 | 64.0 |
|  | c/a | 1.011 | 1.054 | 1.039 | 1.029 | 1.015 | 1.010 |
|  | Ps | 0.24 | 0.47 | 0.41 | 0.35 | 0.28 | 0.27 |
|  | $\Delta E$ | 5.0 | 56.1 | 53.8 | 25.1 | 24 | - |
| PbTiO3 | Eg | 1.47 | 1.88 | 3.00 | 2.08 | 2.83 | 3.60 |
|  | V | 60.4 | 70.4 | 65.2 | 64.9 | 62.4 | 62.6 |
|  | c/a | 1.045 | 1.239 | 1.158 | 1.122 | 1.097 | 1.071 |
|  | Ps | 0.80 | 1.26 | 1.14 | 1.06 | 1.03 | 0.5~1.00 |
|  | $\Delta E$ | 58.1 | 204.8 | 194.1 | 122.7 | 110.6 | - |
| BiFeO3 | Eg | 0.34 | 1.05 (1.76) | 3.4 | 1.89 | 3.0 | 2.74 |
|  | V | 345.1 | 382.7 (384.8) | 375.1 | 369.8 | 369.0 | 373.9 |
|  | Ps | 0.989 | 1.048 (1.003) | 1.103 | 1.027 | - | 1.0 |
|  | $\mu$ | 3.27 | 3.70 (3.95) | 4.1 | 3.96 | 4.2 | 3.75 |